# A Thousand Invisible Cords Binding Astronomy and High-Energy Physics

Rocky Kolb
Department of Astronomy & Astrophysics,
The Enrico Fermi Institute, and
The Kavli Institute for Cosmological Physics
The University of Chicago

> When we try to pick out anything by itself we find that it is bound fast, by a thousand invisible cords that cannot be broken, to everything in the Universe.
>
> John Muir[1]

**The traditional realm of astronomy is the observation and study of the largest objects in the Universe, while the traditional domain of high-energy physics is the study of the smallest things in nature. But these two sciences concerned with opposite ends of the size spectrum are, in Muir's words, bound fast by a thousand invisible cords that cannot be broken. In this essay I propose that collaborations of astronomers and high-energy physicists on common problems are beneficial for both fields, and that both astronomy and high-energy physics can advance by this close and still growing relationship. Dark matter and dark energy are two of the binding cords I will use to illustrate how collaborations of astronomers and high-energy physicists on large astronomical projects can be good for astronomy, and how discoveries in astronomy can guide high-energy physicists in their quest for understanding nature on the smallest scales. Of course, the fields have some different intellectual and collaborative traditions, neither of which is ideal. The cultures of the different fields cannot be judged to be right or wrong; they either work or they don't. When astronomers and high-energy physicists work together, the binding cords can either encourage or choke creativity. The challenge facing the astronomy and high-energy physics communities is to adopt the best traditions of both fields. It is up to us to choose wisely.**

## Introduction

"Astronomy is written for astronomers," Copernicus famously proclaims in the prefatory material of *De Revolutionibus*.[2] We have come a long way since 1543. Now we know that astronomy is not just by and for astronomers. Astronomical discoveries have had a profound impact on many fields of physics and chemistry (and in the future, perhaps biology). They also can be appreciated by an educated populace. Likewise, great advances in astronomy have resulted from the input of many fields of science. One can scarcely imagine astronomy without the physics of Newton's laws, spectroscopy, atomic physics, relativity, quantum mechanics, plasma physics, or nuclear physics.  In turn, astronomy provides a laboratory to study the laws of physics in regimes not easily accessible in terrestrial laboratories.   Astronomy, once a subfield of mathematics, is now more closely allied with the physical sciences, particularly physics.

Since the last decades of the $20^{th}$ century, high-energy physics must be included in the list of fields that have strong binding cords with astronomy.  This is not to say that every field of astronomy has been affected by high-energy physics, just as not every field of astronomy has been affected by nuclear physics or by plasma physics.  But high-energy physics has had an enormous impact on basic questions such as how galaxies and large-scale structure formed and evolved and how the Universe we observe arose from a hot primordial soup of elementary particles in a big bang.

High-energy physics studies the microworld, far removed from our everyday experiences.  The world of neutrinos, quarks, Higgs particles, quantum chromodynamics, or string theory may seem remote to most people.  But, in addition to satisfying humanity's basic curiosity about the workings of the natural world, high-energy physicists invest time, money, and effort studying the laws of the microworld because the laws of nature at the smallest scale determine how objects interact and form the Universe we see about us.

The 'invisible' cords binding astronomy and high-energy physics are quite visible to those who are willing to look.

Of course, it is not just nature on the scales studied by high-energy physicists that determine the behavior of the world about us.  Collective behavior, chaotic behavior, plasma physics, and other macroscopic phenomena play determining roles at various scales of investigation.  Which laws of nature or which particles and interactions are 'fundamental' depends on the scales of the phenomena studied.   In a very real sense, to say that one field of science is more fundamental than another is counter to Muir's vision of a thousand invisible cords binding all of nature.  A division of nature into 'fundamental' sciences and non-fundamental 'generalist' sciences is neither correct nor useful.  If string theory proves to be the correct description of nature, does that mean that high-energy physicists studying electroweak physics or QCD are generalists and not fundamentalists?

One could write an essay about the other cords binding astronomy and different fields of physics or chemistry; plasma physics and astronomy, or general relativity and astronomy, for example.  But here I will discuss the connections between astronomy and high-energy physics.

# On the Nature of High-Energy Experiments and Astronomical Observations

For over 20 years I have been associated with both a high-energy physics (HEP) national laboratory (Fermi National Accelerator Laboratory) and an astronomy and astrophysics department (at The University of Chicago).  During this time I have had a lot of opportunity at these great institutions to observe astronomers and experimentalists at work and play.   It may be surprising to some, but I have always found that astronomers and experimentalists are cut from the same cloth.  I just do not see any real differences.  Both HEP experimentalists and astronomical observers have eyes, hands, organs, dimensions, affections and passions.  But more than that, they are driven by the same resolve to discern the workings of nature.  Galileo would be surprised to learn that astronomers are generalists and physicists are fundamentalists.  Was Galileo a fundamentalist by day experimenting with inclined planes and a generalist by night looking through his optic tube?  Similarities between observers and high-energy physicists are much, much greater than any superficial differences.   However, there are some cultural differences in the way they practice science that should be addressed in collaborations between astronomers and physicists.

*Observational* astronomy is often contrasted to *experimental* physics.  Astronomers obtain information about remote systems (stars, galaxies, etc.,) and cannot in any way control the observational setup or modify any aspect of the source.  The closest they can come to changing external factors is to observe many objects of the same type or observe one object at many wavelengths.  Astronomers studying an object must be patient and await the messenger photons, cosmic rays, or gravitational waves to come to them.  They have no control over the sources they study.

Experiments, on the other hand, study a local system in a controlled environment, and the experimentalist can change external factors like temperature, energy, luminosity, and so on.  Unlike an observer hoping to catch the fleeting afterglow of a gamma-ray burst, a high-energy physics experimentalist can turn their source on or off (at least if the accelerator operators oblige).

Much is made of the differences between observations and experiments by philosophers of science.  But here I will not deal with this distinction.  I believe it is more of a distraction, rather than a distinction.  Perhaps a more significant difference between the methodologies of astronomical observations and HEP experiments is the size and nature of the experimental/observational collaborations.

Let me turn to some of those differences that are important and might be a problem in corroborative efforts at the interface of astronomy and HEP.  They are 1) the size of collaborations, 2) assessing individual contributions for the purposes of hiring and promotions, 3) the nature of astronomical user facilities versus HEP collaborations, 4) and the public release of data to the scientific community.

Astronomy is in many ways more diverse than experimental HEP. Space astronomy, radio astronomy, cosmic-ray astronomy, optical astronomy, high-energy astronomy, and so forth, all have traditions of their own. Here, for the most part, when I speak of observational astronomy, I refer to terrestrial optical astronomy.

## *One Size Does Not Fit All*

Modern high-energy experiments are undertaken by very large collaborations. For instance, ATLAS, one of the two large, general-purpose detectors at the Large Hadron Collider (LHC) at CERN, is a collaboration of 1900 physicists from 164 institutions drawn from 35 countries on 6 out of 7 continents (Antarctica will only be represented by the study of things called "penguin diagrams.") There are three other experiments at the LHC; the collaborations are only marginally smaller. Even the smallest high-energy experimental collaboration is huge compared to *traditional* terrestrial astronomy collaborations.

High-energy physics collaborations were not always this large. The size of the collaborations evolved as the complexity, time-to-completion, and cost of experiments increased. HEP experimentalists work in very large collaborations because that is what is required by the science. Driven to work in large groups, high-energy physicists developed new working models. As with any bureaucracy, the inner workings of the collaboration require things like collaboration councils, collaboration meetings, spokespersons, bylaws, speaker bureaus, conveners, physics coordinators, a structured vetting procedure for publications, and so forth. These seem bewildering to those outside the collaboration (and doubtless to some within). But as evidenced by the continued vitality of HEP, the large collaborations are productive and successful. Also, young, creative scientists still are attracted by the excitement of HEP. They are willing to work in large teams because the science questions are compelling, and large teams are necessary to find the answers to the compelling questions.

Astronomers tend to work in smaller collaborations. This is not to say they work in isolation. Again using terrestrial optical observations as an example, even an observation of a single object by a single astronomer in the course of a single night is the result of an enormous collaborative effort. Modern telescopes are marvels of complex engineering and controls. They were designed and engineered and must be operated in a large collaborative effort. The same is true of state-of-the-art cameras or spectrographs; they are also expensive, complex instruments. While usually not included as authors on the resulting published paper, dozens, perhaps hundreds, of scientists and engineers are behind every observational result.

Astronomy is also evolving towards team efforts. Astronomy surveys of many objects require the intellectual, technical, and telescope resources of many astronomers.

The growing complexity of the investigations demands larger and larger collaborations. But if it is demanded by the science, astronomers can work successfully in collaborations more akin to HEP than to traditional astronomy. Perhaps the best example of such a success is the Sloan Digital Sky Survey (SDSS).[3]

The original SDSS project was a five-year photometric and spectroscopic survey of approximately 10,000 square degrees of the northern sky. The survey was carried out by a collaboration of about 200 astronomers (and high-energy physicists!) from 13 participating institutions. The SDSS was driven to a large collaboration because it was necessary to accomplish the science. Before first light, the SDSS collaboration developed guidelines and procedures for which will appear on collaboration publications. The details of the rules are not relevant here; the SDSS procedure worked splendidly for SDSS, but need not serve as a general model. The important point is that the SDSS experience is evidence that astronomers are able to work successfully within large collaborations.

In the SDSS, astronomers were influenced by and benefited from the large-collaboration experience of HEP, but they did not blindly or uncritically adopt all aspects of the HEP model of large collaborations.

SDSS is a departure from the tradition of small astronomy collaborations, but it has been incredibly successful. As of January, 2007, there have been over 1376 papers with "SDSS" or "Sloan Survey" in the title; these papers have received over 40,000 citations. By any measure, results from SDSS have been heavily used and heavily cited. Of course, many other observatories operating in the traditional user mode have also had enormous impact. Neither large collaborations nor single investigators are appropriate for everything.

The success of large collaborative astronomy projects does not mean that this is the wave of the future in astronomy. Most of astronomy will continue to be done by small groups. It is up to us working in the field to strike a balance between large collaborative projects done in the HEP mode and smaller investigations done in the traditional manner of astronomy. If adopted in a measured fashion, astronomy in the fashion of HEP presents an opportunity, not a danger, to astronomy.

## *Credit Where Credit Is Due*

A typical publication of the SDSS collaboration has a large number of authors by astronomy standards (but a small number by HEP standards). For instance, the paper describing the technical details of the SDSS survey had 144 authors from 36 institutions.[4] It is a highly cited paper; as of August 2007, the paper has been cited 1,356 times according to the Smithsonian/NASA Astrophysics Data System.

In the case of hiring and promotions, how can people outside the collaboration judge the contribution of any one individual to large collaborative efforts? Does the sheer number of authors dilute individual contributions and recognition? If there is one author per paper, then just adding citations might be a good indication of the impact of that author and might be used as a metric for hiring or promotions. But what should be done when there are 54 authors (or 2,054 authors) on a paper? In the instance of the SDSS technical paper with 144 authors, should an author be "awarded" the full 1,356 citations, or $1,356/144 = 9.4$ citations?

This issue has been with HEP for decades, and it is not an insurmountable problem. Members of the collaboration can judge individual contributions. Intelligent promotion and hiring decisions are made after consulting widely within and outside the collaborations. If any astronomy or physics department hires faculty and staff solely on the calculus of the number of citations, or the number of citations per author, they will end up with the department they deserve.

I am sure that one can find instances, perhaps numerous, of people on widely cited papers with large numbers of collaborators who have had little if any intellectual or technical contribution to the science. There will always be people who will abuse any system; perhaps it is easier in very large collaborations. But in practice, this is an insignificant issue.

It is also misleading to regard a HEP experiment as a monolith. It is rather a (highly sophisticated) assembly of many smaller components. Although it is very difficult, if not impossible, to say that one or two people are responsible for the design, fabrication, or testing of an entire detector, it is often easy to identify individuals as leaders in various components of the experiment. As a criterion for hiring or promotion, it is possible to point to significant intellectual accomplishments in parts of a detector or analysis software. Again, this is a tradition in the HEP community, and there is no reason it could not be done in the astronomy community.

## *Facilities versus Experiments*

There is another fundamental difference between HEP and astronomy. High-energy physics is almost completely government funded, while astronomy has the tradition of a significant public-private partnership. A lot has been written about astronomy's public-private partnership, but here I wish to concentrate on an aspect of the *public* support of HEP experiments and astronomical observations.

HEP has experiments, not facilities. High-energy physicists do not write proposals to use a HEP detector for a day or a week to investigate a particular problem. If they want to use a detector to study some physical phenomenon, they must be a member of the collaboration. It's the only path open to them. HEP has user facilities, but not facilities for individual users in the style of astronomy.

Astronomy has publicly supported user facilities (but not enough of them!). Anyone can write a NASA proposal for a space observation using the Hubble or Spitzer Space Telescopes or a NSF proposal to use one of the NSF ground-based observatories. This is a great resource in astronomy, and one that should be maintained as long as possible.

In addition to the user-based mode, recently astronomy has embarked on large collaborative survey projects more in the HEP tradition than the user-facility model. Again, SDSS is a good example of this, but there are others: the 2dF Galaxy Redshift Survey[5] at the Anglo-Australian Telescope, the 2MASS Survey,[6] the Hi-$z$ Supernova Team,[7] and the Supernova Cosmology Project,[8] to name just a few.

There is room for the dual paths of public-access facilities in the astronomy tradition and astronomical experiments in the HEP tradition. In fact, there is often a synergy between them. Astronomy done in the user mode is enriched by astronomy done in the experimental mode. One informs the other. No one mode should be followed exclusive of the other. It is up to us in the field, working with the funding agencies, to strike the proper balance. We can get this right!

## *All Data to the People!*

So far I have concentrated on the spin-off of the HEP experience for the benefit of astronomy. Now let me turn to one example of how I believe the HEP community can benefit from the experience of the astronomy community.

There is a tradition in astronomy for data taken by large federally funded projects to be made public. This tradition arose from the space-astrophysics community, and now it is a very robust aspect of other areas of astronomy. The public release of data has enabled a tremendous amount of science. Of course, it comes with a cost. A non-negligible fraction of the budget of large projects is used to reduce, process, store, and make the data available to the public in a useable form. But I believe that on the basis of the science it enables, in the end it is money well spent.

This is not the tradition in HEP. The data taken by a HEP experiment remains the property of the collaboration. If it is released at all, it is made public only to the extent necessary for the publications of the collaboration. I have never heard a good reason for this, but I have heard four often-repeated excuses.

Excuse #1: *HEP experiments are just too complicated.* Yes, HEP experiments are very complicated. Just measured by the number of channels or the amount of data, large collider experiments are more complicated than even the largest astronomy projects; but both will generate a great amount of data. Let me illustrate future needs by two large experiments, the CMS detector at the LHC, and the planned Large Synoptic Survey Telescope (LSST). Sometime in 2008, CMS expects to take data at an average rate of about 0.2 GB/s. In 2014, LSST plans to take data at an average rate of 0.3 GB/s. LSST plans data releases to the public; CMS does not. Of course the rate of taking raw data is only one measure of the complexity of the data. In the case of collider experiments only about 1 in 10 million events filter through tiers or triggers, while in astronomy virtually all data are recorded. Also, the raw data has to be reduced for analysis, etc. If large astronomy projects are behind HEP projects in the complexity of the data, they are not far behind. Even if HEP experiments are more complicated, eventually the data must distilled into a form for analysis by members of the collaboration. Why not eventually make this data public? Complexity or sheer size of the data should not be an excuse for making data public in some processed form.

Excuse #2: *People would write wrong papers because they don't understand the data.* Yes, I agree completely. But so what? People (at least experimentalists) who continually write wrong papers are eventually marginalized by the community. Science is a self-correcting enterprise. Effort would be wasted chasing down and correcting the wrong papers enabled by public release of data. But the effort would be worthwhile if the release of the data to the public also enables some interesting discovery to be made. Yes, people will write wrong papers; but they will also write correct papers that advance the field.

Excuse #3: *It would be too expensive.* Yes, it would be expensive. For a fixed budget for a detector, money used to develop the end product of a database usable by the general community would mean less money for other aspects of the detector. But would it be worth it? Our goal is to enable discovery and advance science. Perhaps not every astronomer would agree, but I believe that in astronomy money invested in public databases, even at the expense of a less powerful telescope or detector, results in more science, not less science.

Excuse #4: *Members of the collaboration spent a lot of time, effort, and money in producing the data, and it wouldn't be fair if it was made public.* This is not a problem either. In astronomy there is typically a propriety period to allow the people who wrote the proposal, built the equipment, or took the data to have the first opportunity to publish results. The fact that the data will eventually be public has not prevented astronomers from taking the data.

With few exceptions, public data archives are part of the culture of astronomy, but not the culture of HEP. I believe it should be a HEP tradition. Not out of fairness, but because in the long run the experience in astronomy shows that it advances the science.

# Dark Energy and Dark Matter

Now I would like to turn to two examples of the cords binding HEP and astronomy: dark energy and dark matter. I will argue that investigations of dark energy and dark matter by astronomers, high-energy physicists, and by collaborative projects involving both communities, have been mutually beneficial, and if they are done in the right way, will continue to be so.

Over about the last thirty years cosmologists have developed a "standard model" of cosmology. The standard model is a relativistic cosmology (based on Einstein's theory of gravity) of a big-bang expansion from an initially hot and dense state. The remarkable feature of this model is that, in principle at least, it seems capable of explaining all cosmological observations: the character of the cosmic microwave background radiation, the evolution of and the present large-scale structure of the Universe, the abundances of the light elements, and the expansion history of the Universe.

Much has been written about the successes of the standard model of cosmology, but celebrations of its successes should be tempered by the fact that it is based upon unknown physics. In particular, in the standard cosmological model 95% of the mass-energy density of the present Universe is dark; about 25% in the form of dark matter and about 70% in the form of dark energy. Perhaps future cosmologists will look back on dark matter and dark energy as cosmic epicycles of early 21$^{st}$-century cosmology, but most cosmologists today regard them as real phenomena. If they are real, then there is an opportunity in the very near future to discover the nature of the basic mass-energy content of the Universe. It is this opportunity that has attracted the attention of traditional astronomers and high-energy physicists.

## *Dark Energy*

Perhaps the most fundamental quantity in cosmology is the expansion rate of the Universe. In the standard cosmological model the expansion rate depends on the matter-energy content of the Universe, as well as the spatial curvature. In 1997, using Type-Ia supernovae as standard candles, two experimental groups reported evidence for an acceleration of the expansion of the Universe.[7,8] In the framework of the standard cosmological model, an acceleration of the expansion of the Universe requires the mass-energy of the Universe to be dominated by a fluid with a negative pressure. The simplest candidate for this negative-pressure fluid is Einstein's cosmological constant, $\Lambda$.

A cosmological constant is equivalent to, and indistinguishable from, a vacuum energy. If the observations are explained by an effective cosmological constant, then astronomers have made the remarkable discovery that there is a fundamental energy density to the vacuum of space: $\rho_\Lambda \simeq 10^{-30}$ g cm$^{-3}$.

Explaining the small, but non-zero, value of the dark-energy density is a problem that has attracted a great deal of work by string theorists, particle physicists, and cosmologists. What theorist can resist a problem where the naïve estimate is 120 orders of magnitude greater than the observed value? One of the fundamental questions is whether the dark-energy density is indeed "constant" in time, or if it evolves in some dynamical manner as the Universe expands. If we knew the answer to this question it would be an enormous help in finding the correct path toward understanding this important problem.

Unlike dark matter (discussed below), it is likely that the *only* effect of dark energy is to modify the expansion rate of the Universe. If so, then a basic quantity for physicists can only be studied by astronomers! This would be a very visible cord binding astronomy and high-energy physics.

At present, it is fair to say there is no simple, compelling, or elegant solution to explain the observations of the time evolution of the expansion rate of the Universe. Theorists are stumped. Perhaps theoretical physicists should turn to astronomers and use the famous quote of Einstein, "Nothing more can be done by the theorists. In this matter it is only you, the astronomers, who can perform a simply invaluable service to theoretical physics."[9]

The invaluable service to theoretical physics would require a program of large, expensive astronomical projects to determine as accurately as possible the time evolution of the expansion rate of the Universe.

Dark-energy observational programs are a reasonably new thing, but they have attracted a lot of attention and interest. But it should be remembered that only a very small fraction of the astronomy (or HEP) community is actively involved. And although proposed dark-energy experiments are very expensive, there is little danger they will overwhelm the traditional astronomy program.

Furthermore, an intelligently formulated and executed dark-energy program will enable a lot of astronomy. If the program is well executed, experiments will be done, data will be taken, dark-energy information extracted, *and* a lot of astronomy unrelated to dark energy or dark matter will invariably result.

The paranoia that some new program will suck all the oxygen out of the room asphyxiating other programs is not new. There are many examples to point to in the past that have proven baseless. The Hubble Space Telescope (HST) is just one example. There is near unanimous agreement that HST has been great for astronomy. Not only has it led to great advances in astronomy, but it captured the public imagination, inspired students, and attracted many of the best and brightest of the students to pursue careers in astronomy. But before the launch of HST, many voiced concerns that such an expensive telescope would hurt astronomy by diverting interest and resources from traditional ground-based astronomy. Clearly that has not occurred. Not only do we have a thriving ground-based observational community, but a lot of great astronomy is enabled by combining HST and terrestrial observational programs. Plus, it is wrong to assume that astronomy is a zero-sum game; new collaborative community efforts often bring in new sources of funding. (Of course it is also naïve to assume that every new collaborative effort will magically generate new money.)

An intelligently designed dark-energy program will attack astronomical systematic errors to the benefit of other astronomy programs. For instance, in the U.S. there was a Dark Energy Task Force[10] established by the Astronomy and Astrophysics Advisory Committee and the High Energy Physics Advisory Panel to advise the U.S. National Science Foundation, the National Aeronautics and Space Administration, and the Department of Energy on the future of dark-energy research. One of the recommendations of the DETF was for near-term funding of projects that will improve our understanding of astronomical systematic effects. In particular, the DETF identified seven possible projects along these lines:

1. Improve knowledge of the precision and reliability attainable from near-infrared and visible photometric redshifts for both galaxies and supernovae, through statistically significant samples of spectroscopic measurements for a wide range in redshift.
2. Demonstrate weak-lensing observations with low shear-measurement errors and develop lensing methodology and testing on large volumes of real and simulated image data.
3. Obtain high-precision spectra and light curves of a large ensemble of Type Ia SNe in the ultraviolet/visible/near-infrared to constrain, for example, systematic effects due to reddening, metallicity, evolution, and photometric/spectroscopic calibrations.
4. Establish a high-precision photometric and spectrophotometric calibration system in the ultraviolet, visible, and near-infrared.
5. Obtain better estimates of the galaxy population that would be detectable in 21 cm radiation at high redshifts ($2 > z > 0.5$).
6. Develop a better characterization of cluster mass-observable relations through joint x-ray, SZ, and weak lensing studies and also via numerical simulations including the effects of cooling, star-formation, and active galactic nuclei.
7. Support theoretical work on non-linear gravitational growth and its impacts on baryon acoustic oscillation measurements, weak lensing error statistics, cluster mass observables, simulations, and development of analysis techniques.

This list illustrates the cords that bind dark-energy studies to the rest of astronomy. If a dark-energy program helps in any of the above areas, there will be a large beneficial impact on many other areas of astronomy.

## *Dark Matter*

A cosmological term has been in and (mostly) out of favor among cosmologists since Einstein first introduced it in 1917.[11] Dark matter has also been around quite a while. (The history of the dark matter issue can be found elsewhere.)[12] In the 1930's Fritz Zwicky pointed out that unseen (dark) matter was necessary to explain the dynamics of galaxy clusters. In the early-1970's Vera Rubin and Kent Ford established the similar need for dark matter on galactic scales when they measured the optical rotation curves of nearby spiral galaxies.

By the late-1970's, mainstream astronomers accepted the idea that most of the matter in the Universe is dark. Around this time, high-energy physicists became aware of the possibility that in astronomers' quest for accuracy and precision, they may have found evidence for a new type of elementary particle. High-energy physicists at the time could have once again quoted Einstein and said, "How helpful to us in astronomy's pedantic accuracy, which I used to secretly ridicule."[13]

The possibility that the dark matter holding together galaxies and galaxy clusters might be some yet to be discovered elementary particle is a wonderful example of the cords that bind astronomy and high-energy physics. High-energy physicists obtain valuable clues in the search for new particles and new physics from astronomical observations. It is necessary for astronomers to evoke dark matter and dark energy to understand how galaxies and other large-scale structures in the Universe form and evolve. Although large teams of astrophysicists produce enormous numerical simulations of structure formation in the standard cosmological model, without the assumption of dark energy and dark matter the simulations would not agree with observation. Until dark matter and dark energy are understood, they will be little more than epicycles the simulators had to add to save the appearances.

The vitality of the astronomy/HEP interface is well illustrated by the ongoing search for dark matter. If dark matter consists of weakly-interacting massive particles (WIMPs) that were once in thermal equilibrium in the primordial soup, then we could make and study dark matter today if we could re-create the conditions of the primordial soup. The conditions of the primordial soup are today found in the collisions of high-energy beams at accelerators. High-energy physicists are searching through the debris of collisions hoping to find evidence that WIMPs are an ingredient in the primordial soup. If this happens, then for the first time in the last 13.78 thousand-million years WIMPs would have been made, this time as the result of human ingenuity and curiosity! Astronomers are also looking for WIMP signals in the form of high-energy gamma rays, positrons, or antiprotons from present-day dark-matter annihilation in our galaxy, as well as characterizing the distribution of dark matter by measuring the bending of light (gravitational lensing) due to massive clusters of galaxies, determining the motions of stars within galaxies and galaxies within clusters of galaxies, and observing how galaxies and other large-scale structure is assembled. Physicists are also searching underground, undersea, and under ice for the products of dark-matter annihilation in the sun, Earth or the center of our galaxy. Finally there are about a dozen underground experiments hoping to catch the fleeting signal of a relic WIMP passing through a sensitive detector. It is truly a rich experimental program exploring the cords that bind astronomy and high-energy physics.

Dark matter in the form of a new elementary particle is just one possibility. Although now seemingly ruled out by observations, the possibility that the dark matter is in the form of massive astrophysical compact halo objects (MACHOs) was once very popular. Several experimental projects were started to investigate this possibility. The basic idea is that MACHOs would act as gravitations lenses (microlenses in this case), leading to a temporary increase in the apparent brightness of stars in the Large Magellanic Cloud (LMC) as the MACHO crosses near the line of sight to an LMC star.[14]

The microlensing projects are large collaborations by the standards of traditional astronomy. (For instance, there are eighteen MACHO science team members.) The microlensing projects seem to rule out the possibility that the dark matter of our galactic halo consists of MACHOs. Although small in number and much less in cost compared to any of the proposed dark-energy projects, perhaps there is a relevant lesson from the MACHO search experience.

Microlensing projects discovered extrasolar planets—a completely unexpected by-product. Their data also allowed them to use eclipsing binaries for geometrical distance determinations, measure the proper motion of the LMC, and produce a huge database of stars in the LMC. Observations of LMC stars and stars toward the galactic bulge of the Milky Way resulted in a database of 500,000 variable-star observations. The microlensing experiments did not originally plan to learn anything about bulge $\delta$-scuti stars, but they did.

*The lesson is that well planned and well executed experimental projects, even if they lead to null results, can leave behind a rich legacy for astronomy.* This will also happen in the course of an intelligently planned dark-energy program.

## Conclusions

This essay was motivated by the question of whether dark energy is good for astronomy.[15] Although I believe the answer is yes, I also believe that it is the wrong question to ask. A better question is how can the astronomy and high-energy physics communities collaborate and develop a dark-energy program to quantify systematic errors, to investigate theoretical questions about the nature and signatures of different models for dark energy, and finally, to start large observational projects to determine as well as possible the effect of dark energy on the expansion history of the Universe. If the program is conceived and executed in an intelligent way, it will be very good for all of astronomy.  There are right ways and wrong ways to study dark energy.  We are capable of doing it the right way.

Dark energy is an enticing problem.  But there are other, perhaps equally interesting, astronomical questions, such as: are there signs of life on extra-solar planets, what is the nature of dark matter, or did Einstein have the last word on gravity.  One could ask the question whether extra-solar planets are good for astronomy.  That, also, would be the wrong question to ask.   A program to discover and study extra-solar planets is also good for astronomy if it is conceived and executed in an intelligent way.

It is neither useful nor correct to divide science into fundamental and non-fundamental investigations.  High-energy physicists are not fundamentalists and astronomers are not generalists.  They are bound together by a thousand cords that cannot be broken.  Every science is both fundamental and general.

If large collaborations are required to investigate dark-energy science, then astronomers, together with high-energy physicists, can find ways to collaborate without endangering the core activities of either discipline.   If small or individual investigations at user facilities advance science, they will continue to be done.

Astronomy and high-energy physics will benefit from a properly planned and executed dark energy (and dark matter) program.

# Acknowledgments

This essay offers a different view than expressed in Simon White's thought-provoking essay, "Fundamentalist physics: why Dark Energy is bad for Astronomy."[15] I hope it will continue the important discussion Simon's paper began.  I would like to thank the editors for the invitation to contribute a response.  I would also like to thank Dan Green, Kim Griest, Rich Kron, and Tony Tyson for providing information used here.  Finally, I would like to thank many colleagues for discussions about the subject matter of this essay.

---

[1] *The John Muir Papers 1858-1957,* edited by Ronald H. Limbaugh and Kirsten E. Lewis, 1986.

[2] The Latin text in the opening pages of *De Revolutionibus* is "Mathemata mathematicis scribuntur," which is literally translated as "Mathematics is written for mathematicians." At the time of Copernicus, astronomy was regarded as a branch of mathematics (*higher* mathematics, in fact).  This phrase was removed in the "correction" of the book mandated by the Congregation of the Index in 1620.

[3] In the spirit of full disclosure, my institutions are members of SDSS, and I was a member of the SDSS Collaboration Council when publication guidelines were hammered out in the course of countless phone conferences.  Although I once made use of the publicly released SDSS data, I have never been an author on a SDSS paper.

[4] D. York *et al.*, *Astron. J.* **120**, 1579 (2000).

[5] M. Colles *et al.,* astro-ph/0306581.

[6] M.F. Skrutskie, *et al., Astron. J.* **13**, 1163 (2006).

[7] B. Schmidt *et al., Ap. J.* **507**, 46 (1998).

[8] S. Perlmutter *et al., Ap. J.* **517,** 565 (1999).

[9] Einstein wrote this in August 1913 to Berlin astronomer Erwin Freundlich, encouraging him to mount a solar-eclipse expedition to measure the bending of starlight as it passed near the sun. The astronomer eagerly accepted the challenge from the theoretical physicist. Unfortunately for the astronomer, the eclipse of 1914 was in the Crimea during the outbreak of the First World War. In an extraordinary rendition, Freundlich was captured, his equipment confiscated, and he was imprisoned as an enemy combatant. Eventually he was released, but of course he missed the eclipse. This is just as well, because in 1914 Einstein's prediction for the deflection of light by the sun on the basis of his (at the time) incomplete theory of gravity was wrong by a factor of two.

[10] The report of the Dark Energy Task Force is available online at http://www.nsf.gov/mps/ast/detf.jsp. Again, in the spirit of full disclosure, I was a member of the Dark Energy Task Force.

[11] A. Einstein, *Sitzungsbetichte der königliche Preussische Akademie der Wissenschaften zu Berlin*, 142 (1917).

[12] See, e.g., V. Trimble, *Ann. Rev. Astron. Astrophys.,* **25**, 425 (1987).

[13] This is Einstein's statement to Arnold Sommerfeld on December 9, 1915, regarding measurements of the advance of the perihelion of Mercury.

[14] The MACHO collaboration: http://wwwmacho.mcmaster.ca/; The EROS collaboration: http://eros.in2p3.fr; The OGLE collaboration http://www.astrouw.edu.pl/~ogle/.

[15] S. White, *Reports on Progress in Physics.*